**Resonant x-ray scattering investigations of charge density wave and nematic orders in cuprate superconductors**

David G. Hawthorn

Department of Physics and Astronomy, University of Waterloo, 200 University Avenue West, Waterloo, Ontario, Canada, N2L 3G1

email: dhawthor@uwaterloo.ca

## Introduction:

Since their discovery, the "problem" of understanding cuprate superconductors [1] has broadly involved addressing two central questions: what is the microscopic mechanism for superconductivity and how can the superconducting transition temperature be elevated? However, the during the course of over three decades of research superconductors it has been revealed that in cuprates superconductors, and more often than not in other unconventional superconductors, the superconducting phase co-exists and sometimes competes with other types of order, including charge density wave order, nematic order and various forms of magnetic order. Moreover, a route to optimizing superconductivity may entail tuning the relationship between superconductivity and other ordered phases – enhancing or suppressing order (and fluctuations of these orders) in a manner that manipulates the superconductivity. As such, a significant research effort over the decade has focused on identifying ordered phases in unconventional superconductivity and understanding their interrelationship.

Over the past decade, resonant soft x-ray scattering has emerged as an indispensable tool in the effort to investigate intertwined orders in the cuprate and other unconventional superconductors. In particular, resonant x-ray scattering has been widely used to detect and understand the properties of charge density wave and nematic order.

## Resonant x-ray scattering:

Resonant x-ray scattering [2] combines aspects of x-ray diffraction and x-ray absorption spectroscopy. X-ray absorption spectroscopy (XAS) enables investigation of electronic structure. At an absorption edge corresponding to a particular element (ex. the Cu *L* edge), an electron is excited from a tightly bound core state of a particular element into an unoccupied state and the resulting near edge spectrum provides insight into the energies and occupation of the unoccupied states of the element corresponding to the absorption edge. Moreover, using polarized x-rays, the spin and orbital symmetry of the unoccupied states can be identified, leading to linear and circular dichroism of both orbital and magnetic origins. An example of this, shown in figure 1, is the x-ray absorption at the Cu $L_3$ edge of the cuprate superconductor $(La,Nd)_{2-x}Sr_xCuO_4$. These materials famously have a $3d^9$ configuration in the undoped parent compound ($x$ = 0), with the hole in a Cu $3d_{x2-y2}$ orbital. X-ray absorption showed this to be the case: with x-rays polarized parallel to the $CuO_2$ planes, excitation of a $2p_{3/2}$ core electron into the 3$d_{x2-y2}$ orbital is permitted, but not when the x-rays are polarized perpendicular to the $CuO_2$ planes.[3,4]

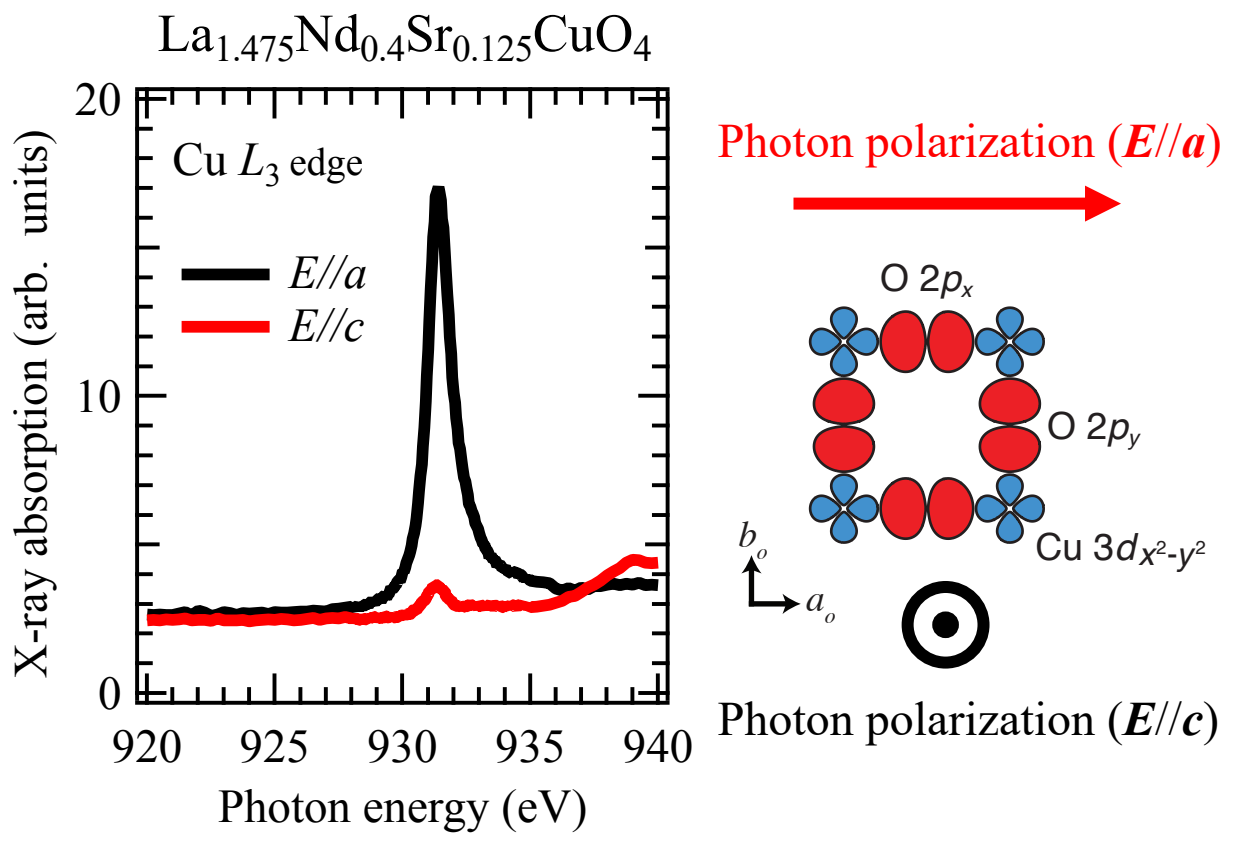

**Figure 1: Polarization dependent x-ray absorption in a cuprate superconductor, $(La,Nd)_{2-x}Sr_xCuO_4$. The XAS with polarization in the $CuO_2$ planes vs perpendicular to the $CuO_2$ planes provides confirmation that cuprates have holes predominantly in 3$d_{x2-y2}$ orbitals.**

This sensitivity to the occupation and symmetry of electronic states carries over to resonant x-ray scattering via an energy and polarization dependence of the x-ray scattering form-factor, $f$. Whereas in conventional Thompson x-ray scattering, $f$ is simply proportional to $Z$, the number of electrons on an atom, on resonance (at or near an absorption edge), $f$ is strongly energy dependent and exhibits the same sensitivity to the energy, occupation and symmetry of unoccupied states as the x-ray absorption. More precisely, for atoms, $j$, in the lattice having the same electronic structure, $f_j(\omega) = f_{j,1} + if_{j,2}$ is a complex quantity, with the imaginary part $Im[f_j] = f_{j,2}$ proportional the atomic absorption cross-section of atoms $j$; $f_{j,2}(\omega) \propto \sigma_j(\omega)$.

An illustration of this is shown in figure 2. Here the real and imaginary components of the x-ray scattering form factor, $f(\omega)$, for Cu is shown over a wide energy range. At high energy, $f(\omega) = 29$, the standard Thompson scattering result. At the Cu $L$ edge, $f(\omega)$ diverges. Zooming in on energies around the Cu $L$ edge reveals sensitivity of $f_j(\omega)$ to orbital occupation. Similar to the x-ray absorption, the energy dependence of the $f_j(\omega)$ will be sensitive to not only valence (the number of unoccupied states per atom), but also to the energies of the unoccupied states relative to the core state of the excited electrons.

Moreover, the natural linear dichroism from orbital polarization or asymmetry, and magnetic linear or circular dichroism from magnetically ordered phases that are evident in x-ray absorption also impact $f_j(\omega)$. Specifically, $f_j(\omega)$ can be expressed as $f_j(\omega, \vec{\varepsilon}, \vec{\varepsilon}') = \vec{\varepsilon}' \cdot F_j(\omega) \cdot \vec{\varepsilon}$, where $\vec{\varepsilon}$ and $\vec{\varepsilon}'$are the incident and scattered photon polarization, respectively, and $F_j(\omega)$ is a rank 2 tensor whose symmetry encodes the local point group symmetry of atoms $j$.[7] It is this sensitivity to the energy, filling and symmetry of the unoccupied states where resonant scattering is most powerful, enabling resonant scattering to probe and differentiate spin, charge and orbital order.

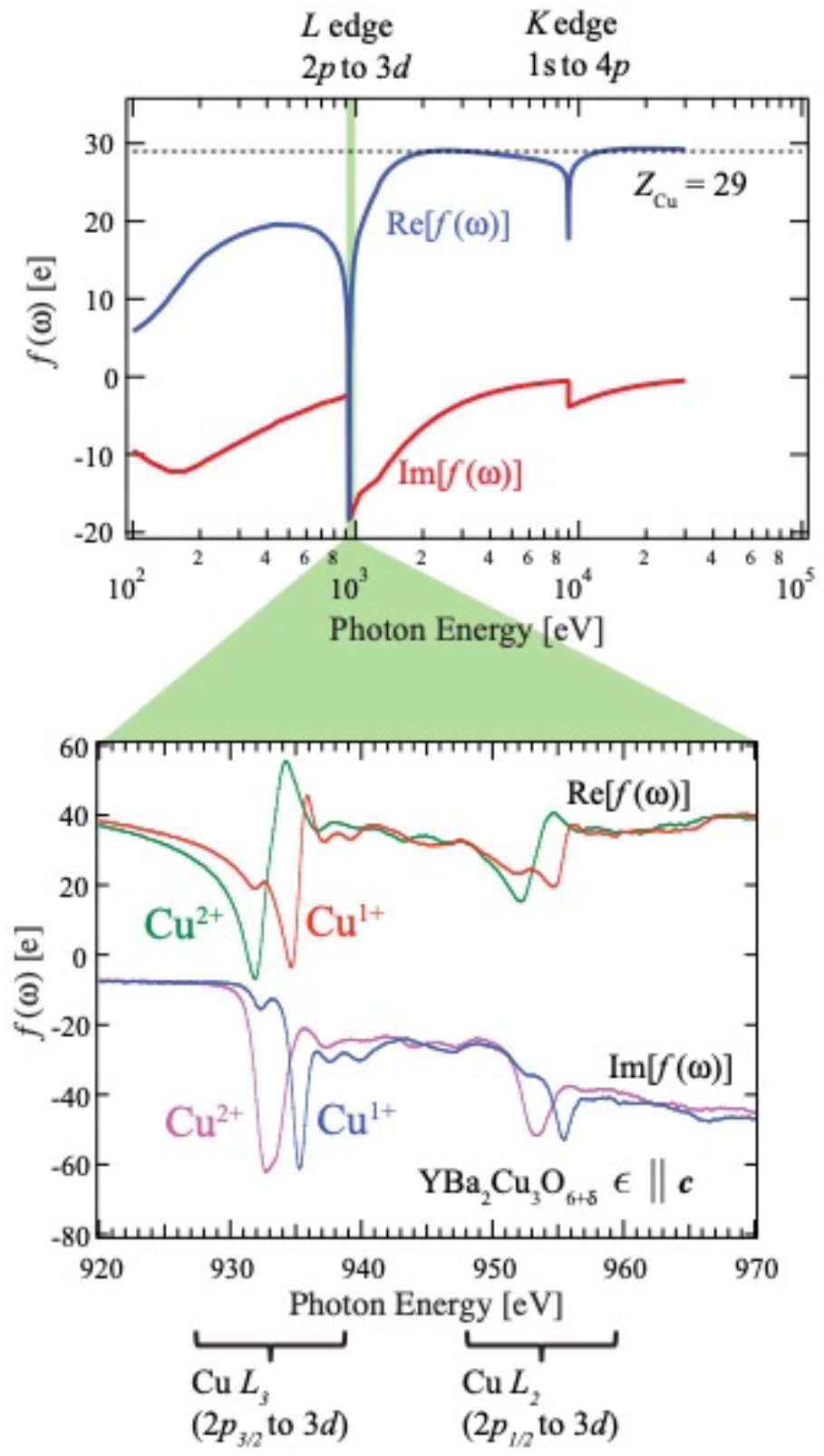

**Figure 2: (Top panel) The tabulated atomic scattering form factor, *f*(ω), from ref. 5 for Cu over a wide energy range. (Bottom panel) The atomic scattering form factor at the Cu *L* edge for Cu sites in $YBa_2Cu_3O_{6+\delta}$ with different valence states. Figure adapted from reference 6.**

The effects of resonant enhancement are most direct when the orbitals accessed in the unoccupied states are states that the very states that are most directly responsible for phenomena like superconductivity, magnetism, charge order or orbital order. In the 3*d* transition metals, these states are typically the 3*d* orbitals. In the rare earths, it is the 4*f* electrons that are most relevant. Moreover, in transition metal oxides, hybridization of the with the O 2*p* states is essential – with many materials, such as the cuprate superconductors being charge-transfer insulators that have doped holes occupying O 2*p* orbitals that are hybridized with Cu $3d_{x2-y2}$ orbitals. These states are all accessed at soft x-rays by measuring the transition metal *L*, rare earth *M* and O *K* edges. Accordingly, the past 2 decades have seen a proliferation of soft x-ray beamlines equipped with in-vacuum diffractors for resonant x-ray scattering from a handful worldwide 20 years ago to dozens of beamlines at present day.

## Detection of Charge Density Wave Order in the Cuprates

A powerful example of resonant scattering has been the study of charge density wave order in the cuprates. The technique was first employed in the cuprates by Abbamonte et al. [8] to measure spin-charge stripes in $La_{1.875}Ba_{0.125}CuO_4$. Although CDW was known to occur in this family of materials, after having been first discovered by neutron scattering via a displacement the lattice induced by charge modulations [9], this study demonstrated the sensitivity of resonant scattering at the Cu *L* and O *K* edges to charge order and provided key insight into the nature of the charge modulation. Specifically, this study verified that the charge order involved a substantive spatial modulation of the Cu 3*d* and in-plane O 2*p* states, but not for instance a modulation of the 2*p* states associated with out-of-plane apical O.

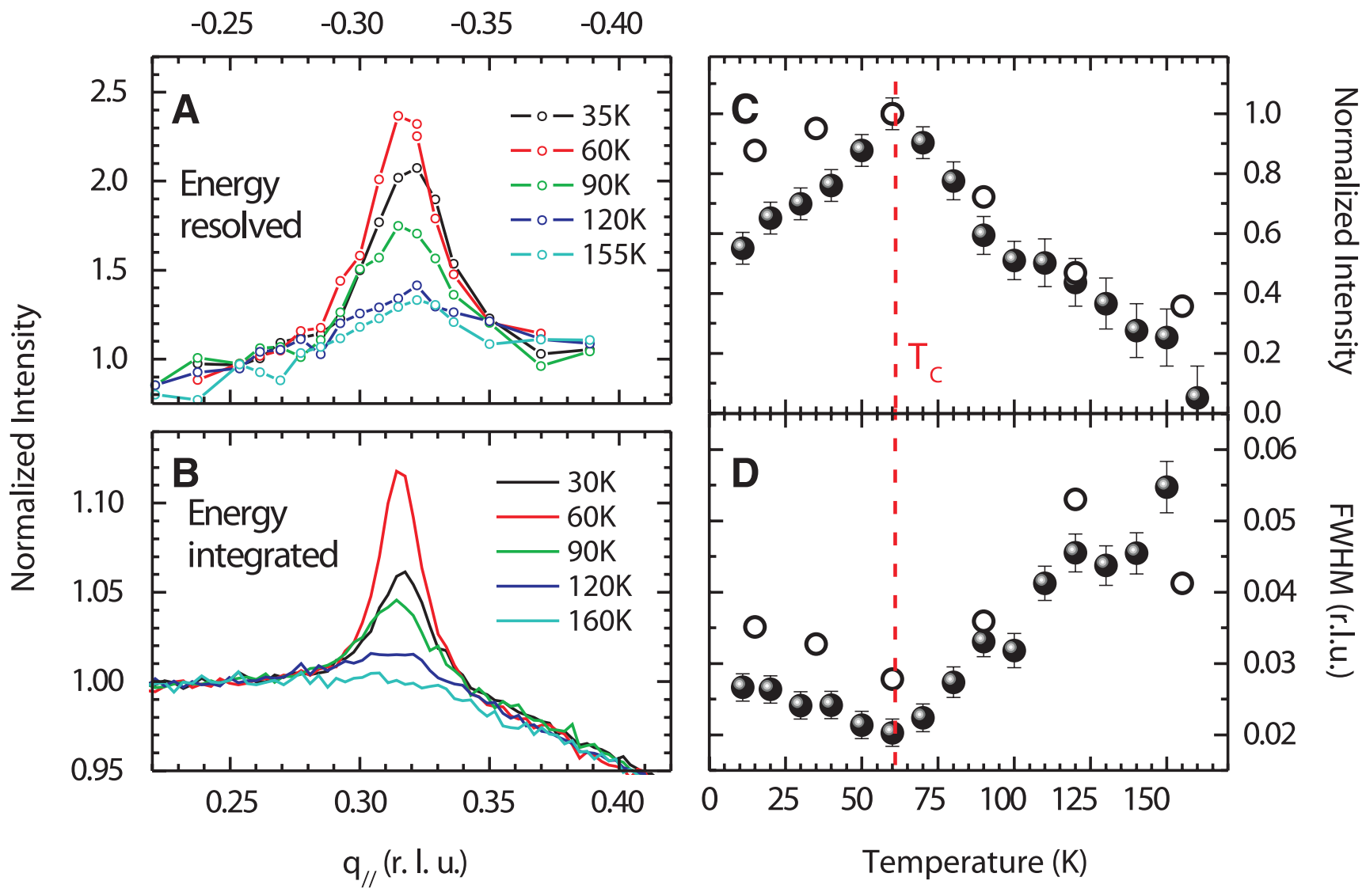

**Figure 3: Resonant scattering measurements at the Cu $L_3$ edge of the CDW Bragg peak in $YBa_2CuO_{6.6}$. The intensity vs in-plane momentum parallel to the CuO bond $q_{//}$ is shown as a function of temperature for both (A) energy resolved (q-RIXS) and (B) energy integrated measurements. In (C), the intensity as a function of temperature is shown, showing an onset around 160 K and a suppression below the superconducting transition temperature, $T_C$. Figure reproduced from ref. 10 with permission.**

The sensitivity of resonant scattering to charge order in the Cu 3*d* planes has been instrumental in discovering charge density wave order in a range of other cuprates. Using resonant x-ray scattering Ghiringhelli et al. provided the first observation of CDW Bragg peaks in the canonical cuprate YBCO [10] (figure 3), after initial observations of related CDW order in high magnetic fields from NMR.[11] This was quickly followed by investigations of CDW in other families of cuprates. Although CDW order had long been observed in Bi2201 and Bi2212 via STM measurements,[12] observation of CDW order via resonant scattering [13,14] that was largely consistent with the STM measurements provided important confirmation from a complementary technique that CDW was indeed present in the materials and moreover was a bulk phenomenon and not simply a surface artifact. Subsequent resonant x-ray scattering measurements provided the first observations of CDW order in Hg-based cuprates,[15] electron doped cuprates,[16] and more recently Tl-based cupates [17]. Due to this battery of resonant x-ray scattering measurements, along with measurements from STM, NMR, neutron scattering and conventional x-ray scattering, it is now evident that CDW order is generic to the cuprate superconductors. Moreover, elucidating the role of CDW order is essential to fulsome understanding the cuprates.

**Microscopic Character of Charge Density Wave Order**

In the above mentioned resonant scattering studies, an important factor in identifying CDW order via resonant x-ray scattering was the enhanced sensitivity to CDW order above a background. Nevertheless, CDW has also been successfully studied using non-resonant hard x-ray scattering in many cuprates,[18,19] indicating that simply using resonant scattering for enhanced contrast is not always necessary.

However, with its sensitivity to electronic structure, resonant scattering has provided additional insight into the microscopic character of CDW order. With non-resonant scattering the CDW order is identified by the displacements of ions that result when CDW order is formed. These studies are also dominated by the displacements of heavy atoms (La, Ba, Sr, …) with only weak sensitivity to the in-plane oxygen atoms that play a central role in the physics of the cuprates. By directly probing spatial modulations of the Cu 3*d* and in-plane O 2*p* states using resonant soft x-ray scattering, greater insight can be gleaned about the microscopic character CDW order, distinguishing CDW order in CuO2 planes from other ordering that may occur.

For example, in the cuprate YBCO, by examining the energy and polarization dependence of resonant scattering, one is able to distinguish a charge density wave in the $3d_{x2-y2}$ orbitals of the Cu(2,3) sites (in the $CuO_2$ planes) from charge modulations in the 3*d* states of the Cu(1) sites in the chain layer (see fig 4a).[20,10] The former are intrinsic to CDW order in the $CuO_2$ planes, but the latter are induced by oxygen dopants ordering into so-called ortho-ordered superstructures. However, CDW order and ortho order can have a very similar period of modulation, close to 3 unit cells and correlation lengths, making them difficult to distinguish. Because $f_{Cu(1),j}(\omega,\vec{\varepsilon},\vec{\varepsilon}')$ differs from $f_{Cu(2,3),j}(\omega,\vec{\varepsilon},\vec{\varepsilon}')$, the energy and polarization dependence of resonant scattering can be used distinguish the two orders. Using x-ray absorption measurements to determine $f_{Cu(1),full}(\omega,\vec{\varepsilon},\vec{\varepsilon}')$, the atomic scattering form factor of Cu in full chain (containing doped O), from $f_{Cu(1),empty}(\omega,\vec{\varepsilon},\vec{\varepsilon}')$, Cu in an empty chain (without doped O), it was possible to model the energy and polarization dependence of the scattering intensity resulting from ortho oxygen order. [6] Agreement between the measured resonant x-ray scattering and the model calculation are excellent in both ortho II and ortho III ordered YBCO. [6,20] In contrast, CDW order in the $CuO_2$ planes is resonant at a lower photon energy and has a characteristically different polarization dependence, as shown in figure 4b. [20]

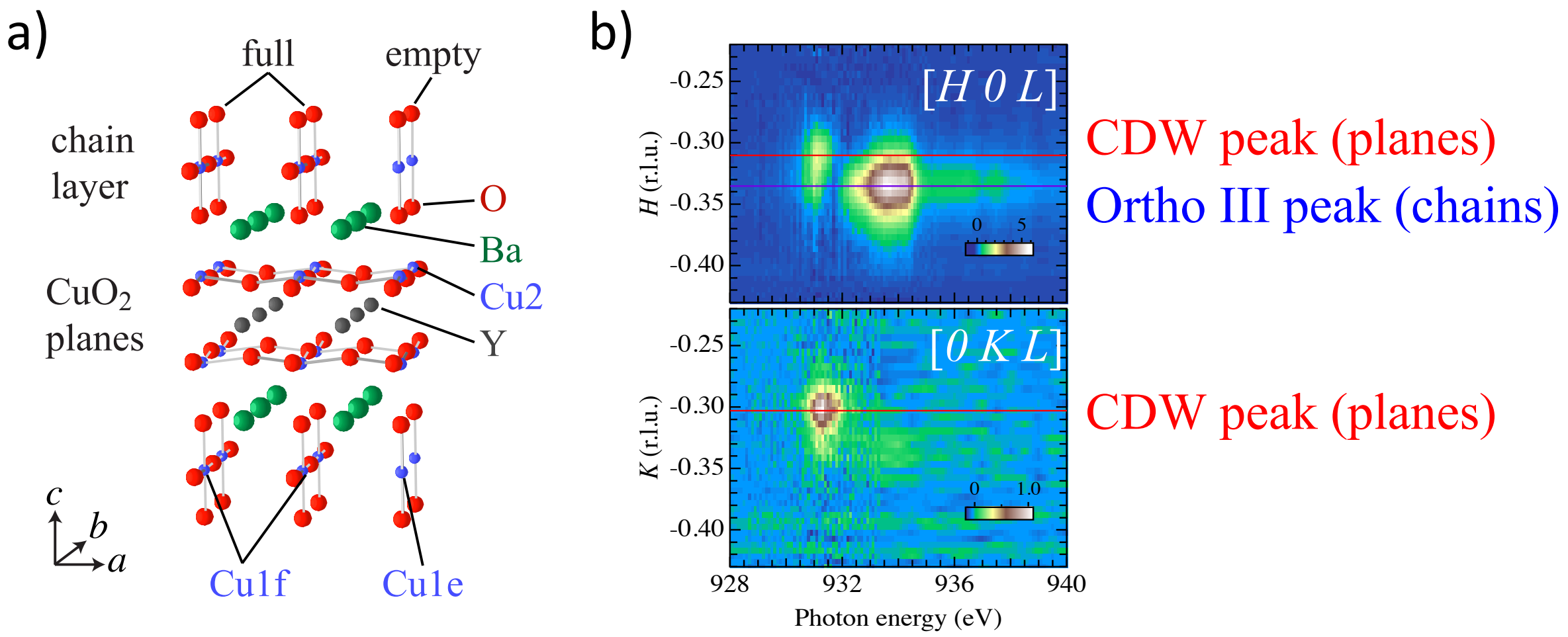

**Figure 4**: **a) Crystal structure of ortho-III $YBCO_{6.75}$. In the ortho-III phase of $YBCO_{6.75}$, the oxygen atoms in the chain layer organize into the full/full/empty pattern shown here along the *a*-axis. Although this leads to charge (or valence) order in the chains, the CDW order in YBCO forms in the $CuO_2$ planes along both the *a* and *b* axes. b) Energy and polarization dependence of resonant scattering at the Cu $L_3$ edge of ortho-III $YBCO_{6.75}$. The resonant scattering intensity (ref 20) is shown for scattering along *H* (*a*-axis, top panel) and *K* (*b*-axis, lower panel) for sigma incident photon polarization. The ortho III chain order peaks only appear in *H* (along *a*) and resonate at higher photon energy (933.6 eV) than the CDW order from the $CuO_2$ planes, which appears in *H* and *K* (along *a* and *b*) at an energy of 931.3 eV. The ortho III and CDW peaks are also separated in *Q*, with the CDW peak at *H*(*K*) = 0.31 and the ortho III peak at *H* = 0.33.**

In addition to provide the contrast needed to identify charge modulations in the $CuO_2$ planes from the ortho ordering, the resonant x-ray scattering has provided unexpected insight into the microscopic character of CDW in the planes.

A simple picture of charge density wave order involves a modulation in the hole density of the O 2*p* states, which are hybridized with the Cu 3*d* states. Particularly in a strongly correlated material such as the cuprates, a modulation in hole density, would be accompanied by a relaxation of the lattice (ionic displacements), as well as, a modulation in the local electronic structure. The latter is parameterized by a range of parameters, including the hopping between Cu 3*d* and O2*p* states, $t_{pd}$, the energy of the Cu 3*d* hole states, $\epsilon_d$, the energy of the O 2*p* states, $\epsilon_p$, the charge transfer energy $\Delta_{pd} = \epsilon_d - \epsilon_p$, and the effective nearest neighbour exchange interactions, $J$. In assessing the microscopic character of CDW order, and its origins, it is unclear which parameters should be considered essential ingredients to describe the CDW phase. Resonant scattering provides an opportunity to differentiate and identify the most dominant contributions (energy shifts, orbital occupation, lattice displacements) contributions to the CDW scattering intensity.

Examining the energy dependence of resonant scattering from CDW order in the cuprates, Achkar et al. [21,20] found that the energy dependence of CDW Bragg peak intensity at the Cu *L* and O *K* edges in $La_{0.1475}Nd_{0.4}Sr_{0.125}CuO_4$ and Cu *L* edge scattering in YBCO is explained well by a very simplified model involving only modulations in the Cu 3*d* or O 2*p* orbital energies, $\epsilon_d$ and $\epsilon_p$ (see figure 5). In contrast, a large modulation of orbital occupation, which is perhaps expected and originally argued by Abbamonte et al. [8] to explain the resonant scattering at the O *K* edge, is not a good match for the measured energy dependence. Although charge density modulations no doubt occur, they appear to have a minimal direct signature in the measured energy dependence of the CDW resonant scattering. This suggests that at a microscopic scale, strong correlations are playing a role in modulating the local electronic structure and orbital energies in the CDW phase.

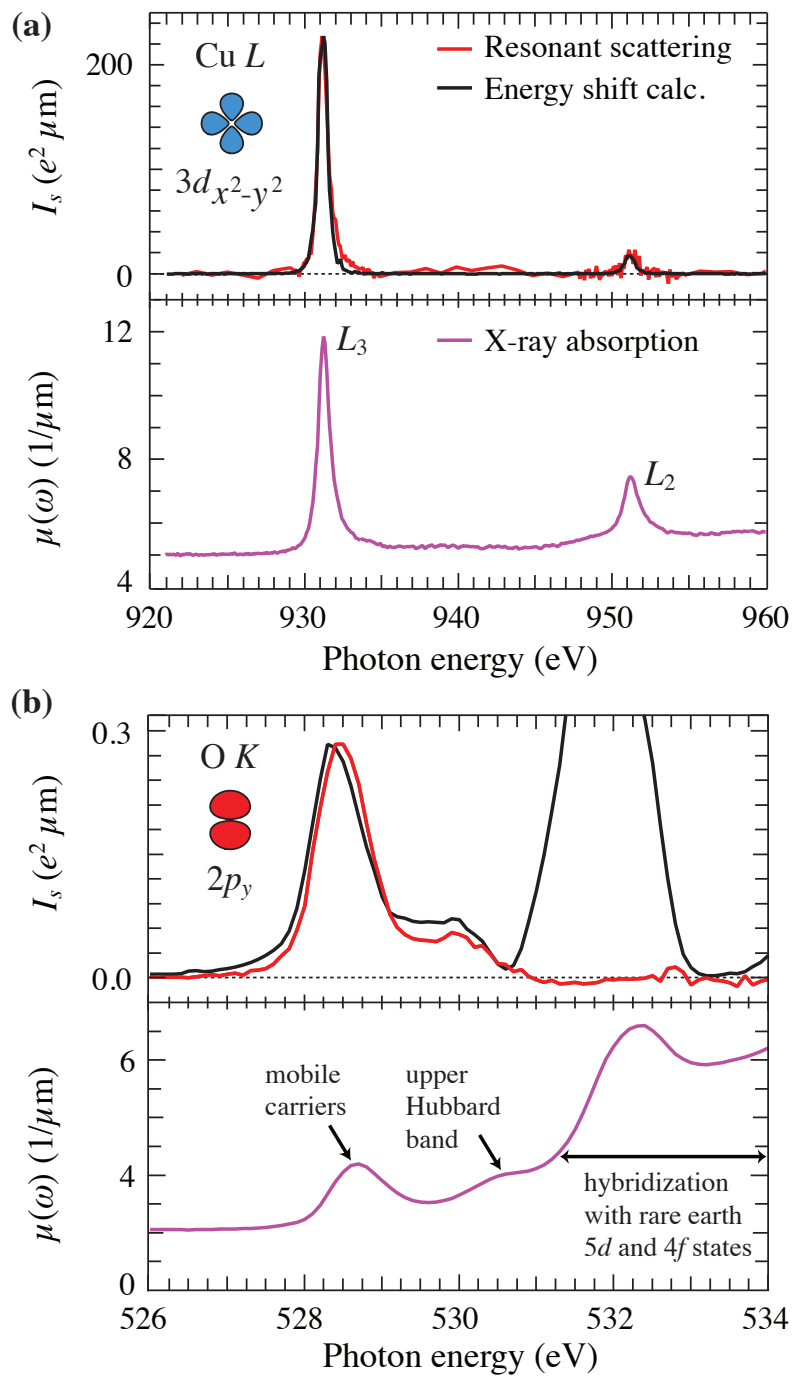

**Figure 5: Resonant scattering of 1/8 doped LNSCO at the Cu *L* and O *K* edges. (Top panels) The measured scattering intensity (red line) is compared to the energy shift model calculation (black line) at the Cu *L* (a) and O *K* (b) edges. (Lower panels) X-ray absorption coefficient (magenta line) measured at the Cu $L_{3,2}$ (a) and O *K* (b) edges using total electron yield. Figure adapted from ref. 21.**

## Orbital Symmetry of CDW order:

The original identification of charge stripes in the cuprates indicated a unidirectional modulation of charge density [9]. In analogy with the smectic and nematic phases of liquid crystals, it was suggested that the CDW phase of may correspond to an electronic liquid crystal, characterized by both a translational symmetry breaking order parameter (akin to the smectic order) that is evidenced by superlattice Bragg peaks, and an Ising-nematic order parameter, indicative of rotational symmetry breaking of the electronic structure from $C_4$ to $C_2$.[22,23] In other words, CDW order is characterized by a $\boldsymbol{Q} = 0$ nematic order, as well as, a $\boldsymbol{Q} = \boldsymbol{Q}_{\boldsymbol{CDW}}$, translational symmetry breaking order.

Moreover, orbital arrangements of the CDW modulations represented by the $\boldsymbol{Q} = \boldsymbol{Q_{CDW}}$ order may be non-trivial. For instance, it is possibility that the charge density modulations also involve orbital modulations, with the energy and occupation of Cu 3*d*, O $2p_x$ and O $2p_y$ orbitals modulated with different amplitudes and phase, potentially leading to forms of charge/orbital order such as a *d*-form factor (or quadrupolar) ordered CDW where O $2p_x$ orbitals are modulated 180 degrees out-of-phase with the O $2p_y$ orbitals. [24, 25,26,27,28] Resonant scattering has been employed to investigate both of the $\boldsymbol{Q} = 0$ and finite $\boldsymbol{Q_{CDW}}$ character of CDW order in the cuprates.

We will consider first the orbital symmetry of CDW order at finite $\boldsymbol{Q} = \boldsymbol{Q_{CDW}}$. The orbitals involved in CDW order can be investigated by measuring the intensity of the CDW Bragg peak while varying the projection of the incident photon polarization onto the crystallographic axes of the sample. This allows sampling of different components of the scattering tensor. Most powerfully, this is done by measuring with both sigma and pi incident photons while rotating the sample azimuthally about $\boldsymbol{Q_{CDW}}$. Comin et al. utilized this approach at Cu *L* edge in YBCO, reporting a signature in the azimuthal dependence of the scattering intensity that was more consistent with a *d*-form factor CDW order than an *s* or *s'* form factor CDW, [29] nominally consistent with findings from STM measurements on Bi2212 and NaCCOC of a *d*-form factor CDW in those compounds.[30,31] However, the contrast in the initial observations from Comin et al. was close to the variance in the data. Follow up measurements in YBCO in a geometry with greater sensitivity to *d* vs *s* form factor CDW order,[32] reached a different conclusion – that the azimuthal angle dependence of the scattering at the Cu *L* edge is YBCO is much more consistent with a *s* form factor CDW order (at least for the [0 *K L*] CDW peak), with no evidence of a *d* form factor CDW order. Notably, this does not rule out a *d* form factor CDW in YBCO, but indicates if this does indeed occur in YBCO, it does not result in a clear signature in resonant scattering at the Cu *L* edge.

More direct measurements of the orbital symmetry of CDW order were performed in $La_{1.875}Ba_{0.125}CuO_4$, using measurements at the O *K* edge, providing direct sensitivity to modulations of the O $2p_x$ and O $2p_y$ orbitals.[33] Here, the CDW order is found to be consistent with a predominantly s' form factor, indicating charge in the O $2p_y$ are principally modulated in phase with each other. However, the O $2p_x$ and O $2p_y$ exhibit an unequal magnitude of modulation, with a 0.6 smaller modulation of the $2p_x$ relative to the $2p_y$ states for CDW order propagating along the *x* direction. Such an asymmetry indicates orbital as well as charge degrees of freedom are needed to describe the CDW state.

A perhaps more striking example of the orbital character of CDW order in the cuprates is in YBCO.[32] Here the CDW order propagating along the *b* axis is consistent with a modulation of $3d_{x2-y2}$ orbitals, with a small (~5%) and expected hybridization with $3d_{3z2-r2}$ states. However, the CDW order propagating along the *a* axis, evidenced by Bragg peaks at [*H* 0 *L*], exhibit a highly non-trivial orbital symmetry described by orbital rotations out the $CuO_2$ planes (tilting of orbitals) or an anomalously large contribution of out-of-plane orbitals ($3d_{3z2-r2}$ ,$3d_{yz}$, $3d_{xz}$) to the CDW order. Although the importance of orbital physics may not be generic to all manifestations of CDW order in the cuprates, these measurements highlight that orbital degrees of freedom (orbital occupation and orbital energy) can be spatially modulated in the CDW phases of the cuprates, perhaps necessitating a description of CDW order in the cuprates as a charge/orbital order in some instances.

**Observation of Nematic Order in the Cuprates from Resonant X-ray Scattering:**

Resonant scattering has also been used to study $\boldsymbol{Q}_{x,y} = 0$ Ising nematic order in the (La,RE)$_2$CuO$_4$ cuprates. [34,35,36] This is done by measuring a specific Bragg peak, the 001 peak, which provides sensitivity on resonance to orbital symmetry instead of ionic positions.

In (La,RE)$_2$CuO$_4$ cuprates, some materials doped with Nd, Eu or Ba, exhibit a low temperature tetragonal (LTT) structural phase that is characterized by CuO$_6$ octahedra that rotate about the the CuO bond direction. This induces an asymmetry in the electronic structure within individual CuO$_2$ planes that has long been associated with the formation of unidirectional spin-charge stripes.[9] However, the octahedral tilt axis rotates by 90 degrees between neighbouring layers, giving a globally tetragonal structure. The 001 Bragg peak on resonance was identified as a means to characterize the transition to this structural phase. [37,38,34,35] Although this peak is forbidden in conventional diffraction (from the perspective of ionic positions along the *c* axis, there is always an identical layer half a unit cell away along *c* – resulting in destructive interference at 001), this peak is allowed on resonance due to the sensitivity of resonant scattering to orbital symmetry, which differs between neighbouring planes. The resulting intensity of the 001 peak is proportional to the difference in the orbital symmetry between planes, corresponding to a well-defined $\boldsymbol{Q}_{x,y} = 0$ Ising nematic order parameter.

Significantly, the 001 peak intensity and its temperature dependence can be measured at photon energies corresponding different atoms in the unit cell, allowing one to separately probe orbital asymmetry in the CuO$_2$ planes relative to the spacer layer. Achkar et al [34] found that the temperature dependence of the 001 Bragg peak differs for atoms in the spacer layer (whose orbital symmetry is expected to closely track the ionic positions), versus the atoms in the CuO$_2$ planes, which exhibited a more gradual temperature dependence. This is interpreted as evidence of a nematic order in the CuO$_2$ planes that is distinct from, but naturally coupled to the symmetry breaking lattice distortions in the spacer layer.

Moreover, Achkar et al [34] found that the intensity of the 001 peak at the Cu *L* and in-plane O *K* energies is enhanced below $T_{CDW}$, the CDW onset temperature, showing that the $\boldsymbol{Q}_{x,y} = 0$ nematic order parameter is enhanced by the $\boldsymbol{Q} = \boldsymbol{Q}_{\boldsymbol{CDW}}$ order parameter.

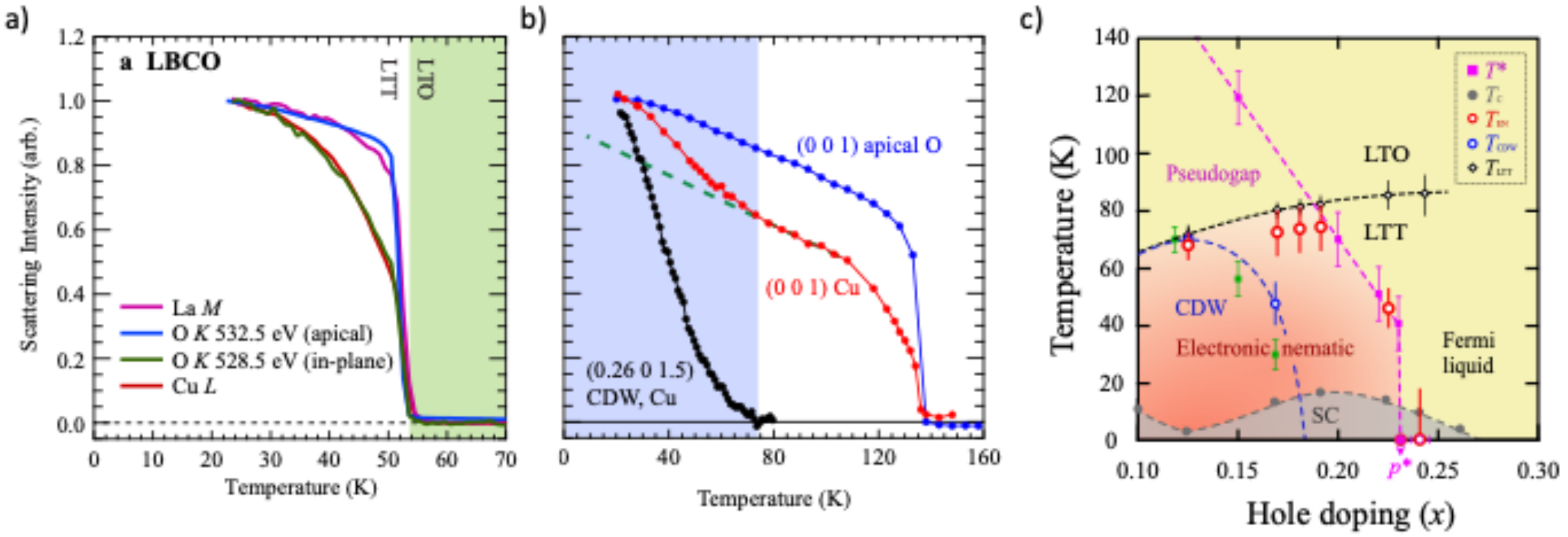

**Figure 6: a) The intensity of the (001) Bragg peak intensity as a function of temperature in $La_{1.875}Ba_{0.125}CuO_4$ at different photon energies from ref. [34]. The different temperature dependence of *I*(001) when probed at photon energies corresponding to the $CuO_2$ planes relative to the La and apical oxygen provides evidence of nematic order. b) The (001) peak intensity vs *T* in In $La_{1.65}Eu_{0.2}Sr_{0.15}CuO_4$, the from ref. [34] showing nematic order is enhanced below $T_{CDW}$, the CDW onset temperature. c) The phase diagram of nematic order in $La_{1.6-x}Nd_{0.4}Sr_xCuO_4$ from ref. [35] indicating that signatures of nematic order persist only in the pseudogap phase and also occur int eh absence of detectable CDW order.**

While the study by Achkar et al. made use of differences in the temperature dependence to measure nematic order, a recent study has explored the dynamics of nematic order on ultra-fast timescales using optical pump- x-ray probe resonant x-ray scattering at the Pohang FEL. [36] Here a fs 800 nm pump laser is used to excite the sample, disrupting the CDW order and nematic order. The response of the 001 Bragg peak is measured with sub-ps time resolution. This study found a pronounced difference in the dynamics of the 001 Bragg peak when measuring at the Cu *L* edge and in-plane O *K* edge relative to the La *M* or apical oxygen *K* edge (the latter probing atoms in the spacer layer). These dynamics confirm an electronic origin to the nematic order, as well as a coupling with the translational symmetry breaking, $\boldsymbol{Q} = \boldsymbol{Q}_{\boldsymbol{CDW}}$. However, being a non-equilibrium probe, these new measurements provide additional insight into how CDW order melts, which is characterized by both an amplitude of a charge modulation and a phase.

Finally, explorations of nematic order in the cuprates have also revealed that, although nematic order appears coupled to CDW order, signatures of nematic order survive at dopings and temperatures where CDW order is not present (or too weak or short range to be detected).[35] Moreover, nematic order appears to be connected to the pseudogap phase, with signatures of nematic order disappear above the pseudogap onset temperature, $T^*$ or doping $p^*$. [35]

Recently STM measurements have shown that orbital order degrees of freedom such as the charge transfer energy have been identified to be important for setting the superconducting gap of cuprates [39] and play a role in forming short range orbital/nematic order in Bi2212 [40]. The modulations in orbital energy in the CDW phase of cuprates [21,20] and observation of nematic order [34,35,36] by resonant scattering are likely probing the same phenomena.

## Conclusions

Many open questions remain regarding the origins of nematic and CDW order in the cuprates, such as how these orders manifest in different cuprate materials with varied crystalline structures and howe these orders are intertwined with superconductivity. Resonant x-ray scattering has proven a valuable tool to attempt to probed CDW and nematic orders in a range of cuprate materials, providing important insight into the orbital character of CDW order.